\documentclass[twocolumn,preprintnumbers,superscriptaddress,landscape,nofootinbib,prl]{revtex4}
\usepackage[utf8]{inputenc}
\usepackage{amsmath}
\usepackage{amsthm}
\usepackage{float}
\usepackage{braket}
\usepackage{graphicx}
\usepackage{url}
\usepackage[colorlinks=true, pdfstartview=FitV, linkcolor=red, citecolor=blue, urlcolor=blue]{hyperref}
\usepackage{slashed}
\usepackage[normalem]{ulem}

\graphicspath{{./figures/}}

\newcommand{\be}{\begin{equation}}      
\newcommand{\ee}{\end{equation}}      
\newcommand{\bea}{\begin{eqnarray}}      
\newcommand{\eea}{\end{eqnarray}}

\begin{document}

\title{Comments on "Minimal Model for Fast Scrambling"}

\author{Kazuki Ikeda}
\email[]{kazuki.ikeda.gt@kyocera.jp}

\affiliation{Research Institute for Advanced Materials and Devices, Kyocera Corporation, Soraku, Kyoto 619-0237, Japan}


\bibliographystyle{unsrt}

\if{
\begin{abstract}
I consider a spin chain with the nearest-neighbor and the next nearest-neighbor interactions. I show that the Hamiltonian obtained by specifying the desired combination of symmetry blocks is a fast scrambler. This is a smaller model than the "minimal model" by  Belyansky et al.~\cite{2020PhRvL.125m0601B}.  
\end{abstract}
}\fi
\maketitle

\begin{figure}[H]
\includegraphics[width=\hsize]{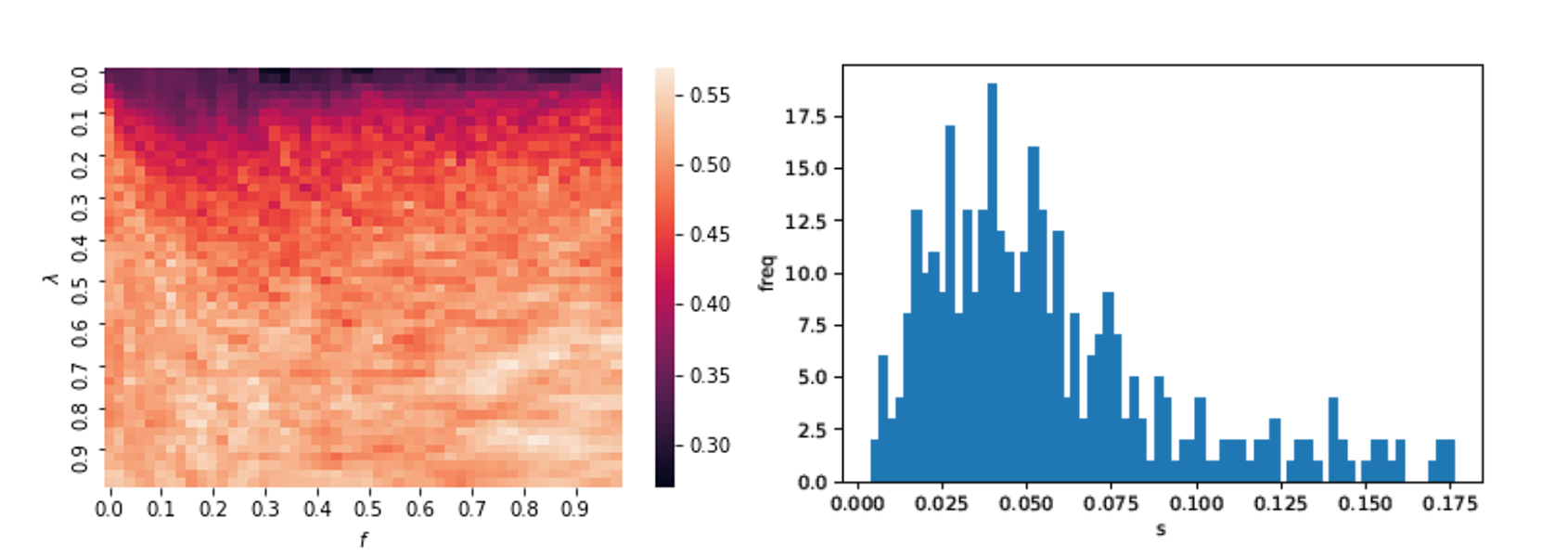}
    \caption{[Left] The average of the $r$-parameter becomes maximal 0.57 at $(\lambda,f)=(0.90,0.84)$, if $g=1$. [Right] the corresponding histogram of $s$. 
    }
    \label{fig:r-params}
\end{figure}

\begin{figure}
\begin{minipage}{0.49\hsize}
    \centering
    \includegraphics[width=\hsize]{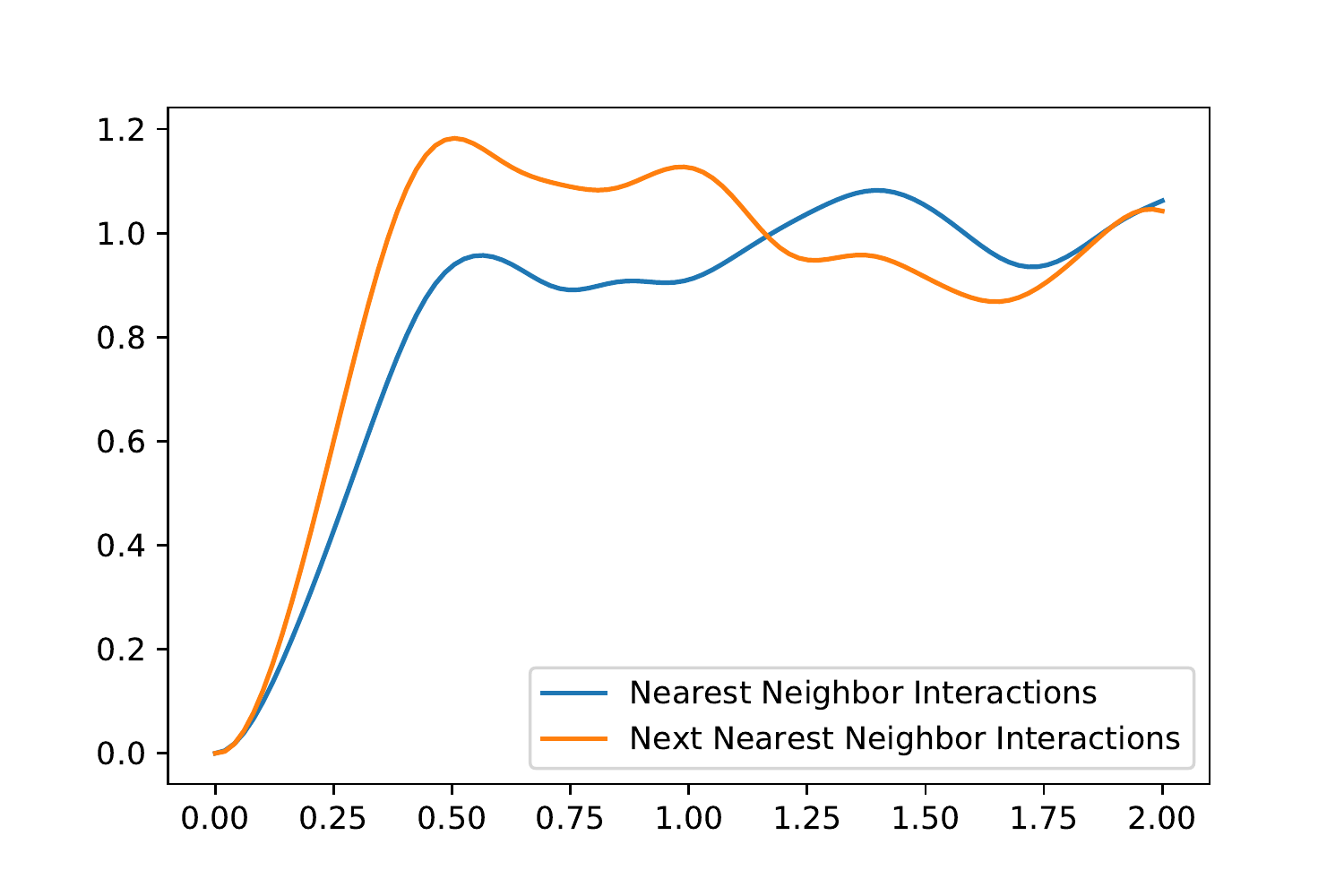}
\end{minipage}
\begin{minipage}{0.49\hsize}
    \centering
    \includegraphics[width=\hsize]{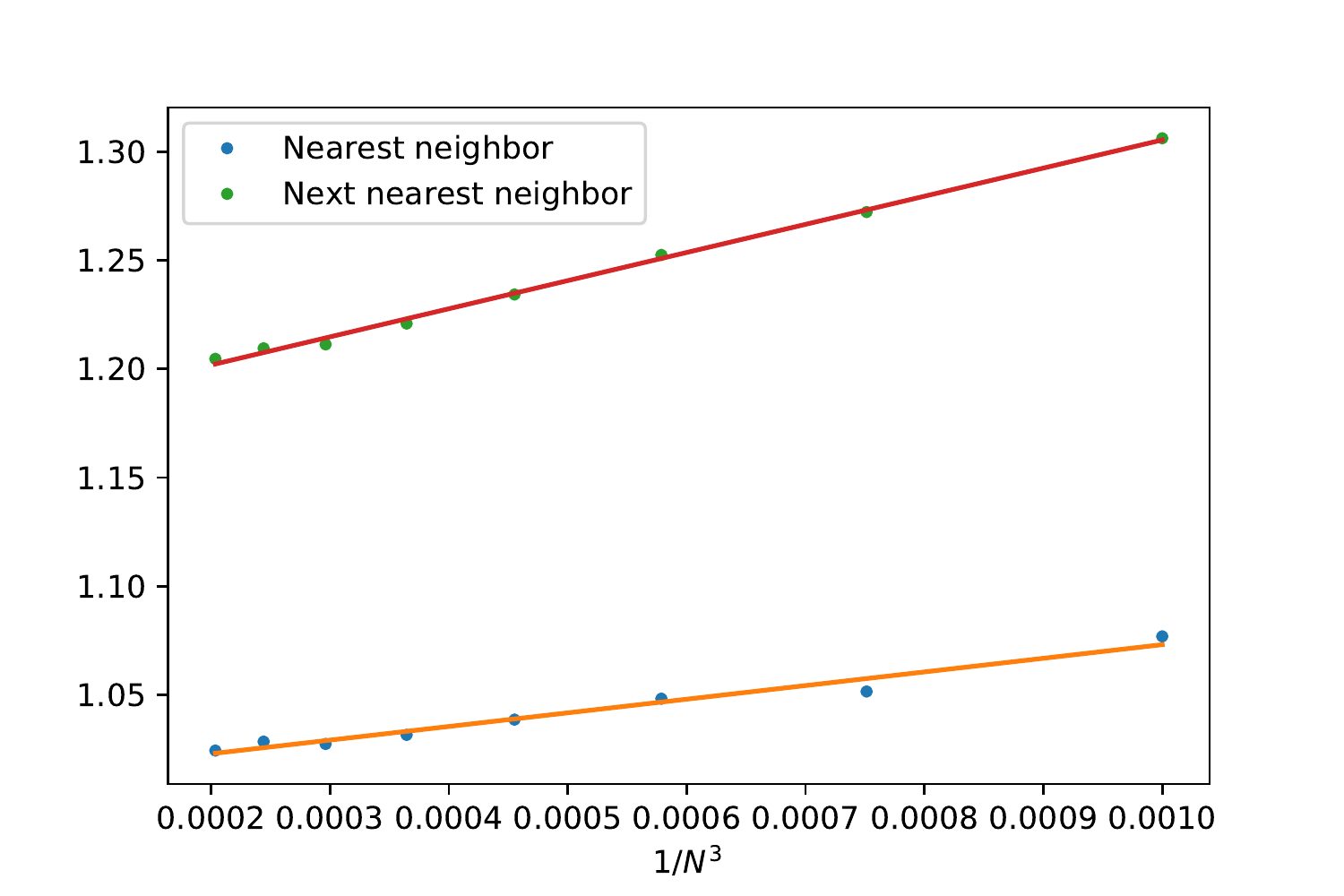}
\end{minipage}
\begin{minipage}{0.49\hsize}
    \centering
    \includegraphics[width=\hsize]{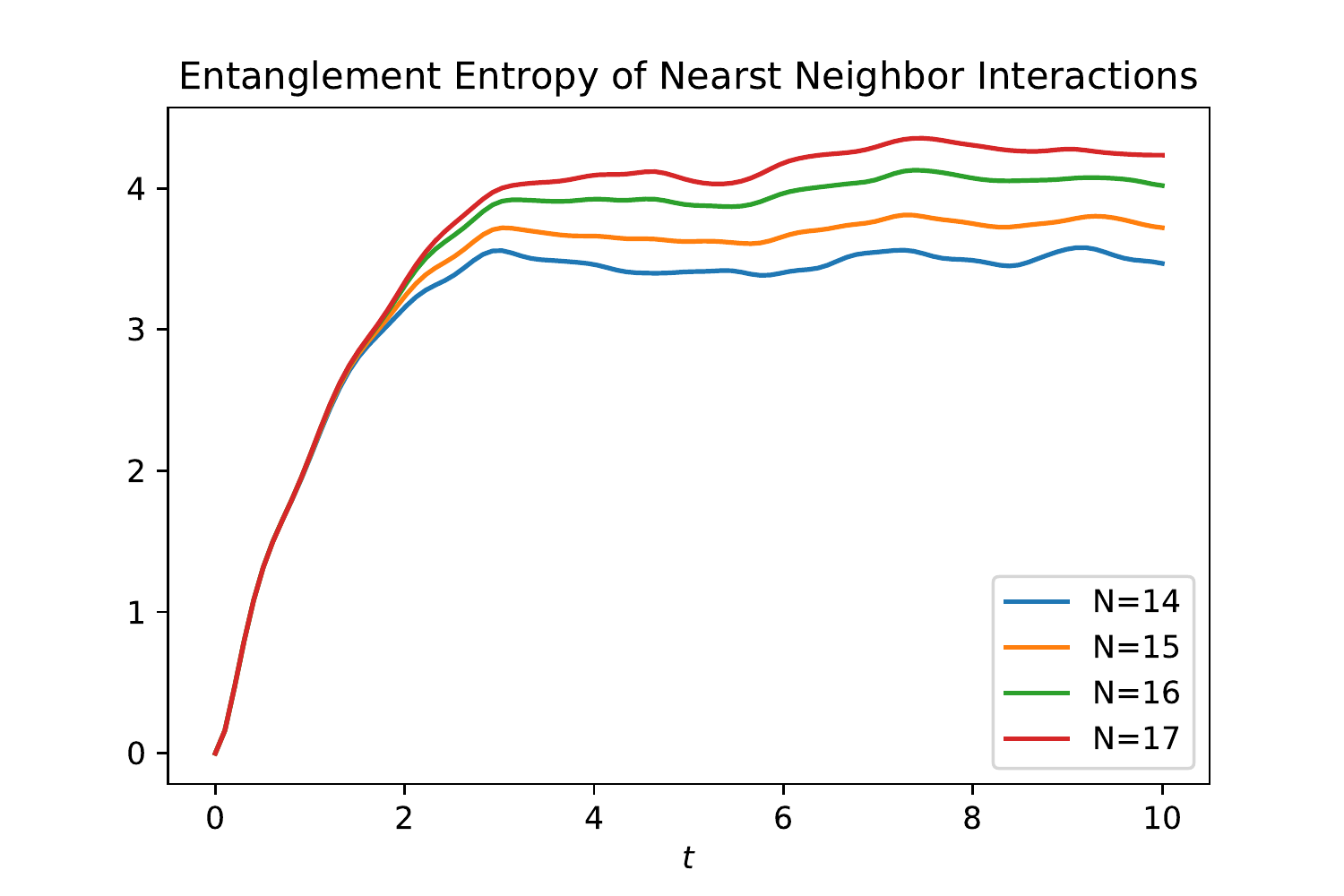}
\end{minipage}
\begin{minipage}{0.49\hsize}
    \centering
    \includegraphics[width=\hsize]{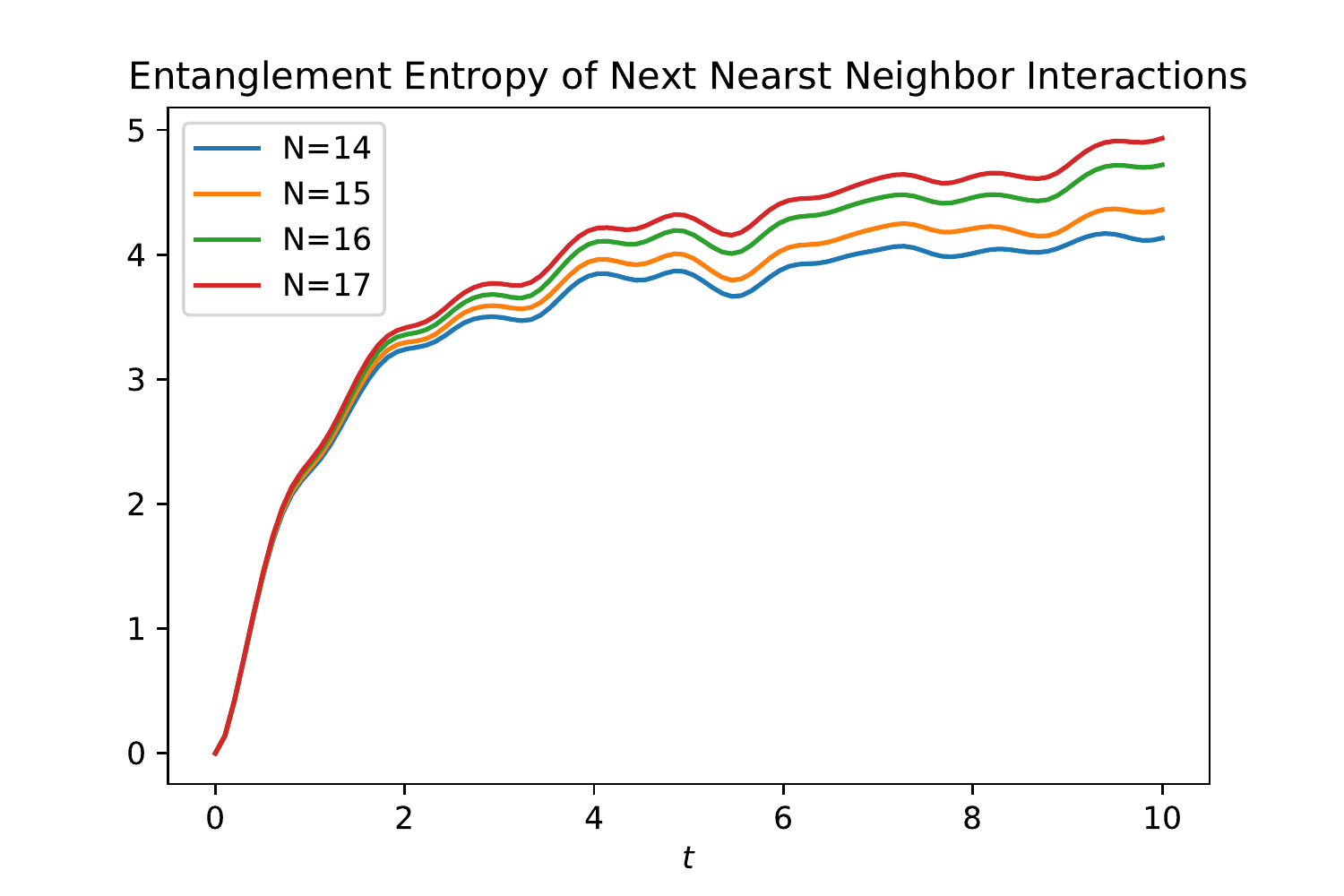}
\end{minipage}
\caption{[Top Left] Time-dependence of $C(t,r=N)$ with $N=17$ spins. [Top Right] Plots of $C(t=t_\ast,r=N)$ up to $N=17$. 
[Bottom] Half-cut entanglement entropy of NN-interactions and NNN-interactions, starting from the paramagnetic state $\bigotimes_i\ket{+}_i$, where $\ket{+}_i=\frac{1}{\sqrt{2}}(\ket{0}_i+\ket{1}_i)$. }
    \label{fig:linear}
\end{figure}

In the Letter~\cite{2020PhRvL.125m0601B} the authors studied a variant of
fast scrambling models. They claim that their all-to-all interacting model is the minimal as a fast scrambler. In this Comment I show that this is not the case. To elaborate on this issue, I show that a Hamiltonian obtained by specifying desired combination of some sectors can be a fast scrambler regardless of the absence of all-to-all interactions. To this end, I consider a spin chain 
\begin{align}
\begin{aligned}
\label{eq:Ham}
    H=&-\frac{1}{1+\lambda}\left(\sum_{i}Z_iZ_{i+1}+\lambda\sum_{i}Z_{i}Z_{i+2}\right)\\
    &-f\sum_{i} X_i-g\sum_{i}Z_i. 
\end{aligned}
\end{align}
I work with the system in the zero-momentum sector of positive parity. This model is non-integrable for general $\lambda,f,g\neq0$. To see this I compute the average $\langle r\rangle$ of the $r$-parameter which is the average of $r_n=\frac{\min\{s_n,s_{n-1}\}}{\max\{s_n,s_{n-1}\}}$ for $s_n=E_n-E_{n-1}>0$.
It is known that $\langle r\rangle $ takes some benchmark values in the case of chaotic systems. For example, $\langle r\rangle _\text{GOE} \approx 0.54$ for random matrices of the Gaussian Orthogonal Ensemble (GOE). My case is exhibited in Fig.\ref{fig:r-params}, where the level spacing distribution shows a Wigner distribution, hence the model is chaotic. In what follows I use $(\lambda,f,g)=(0,1.05,0.5), (0.9,0.84,1)$ for the nearest neighbor interactions (Ising model) and the next nearest neighbor interactions, respectively. 
I consider $C(t,r)=1-\text{Re}[\langle X_1(t)X_rX_1(t)X_r\rangle]$,
where the expectation value is evaluated in a Haar-random pure state. I find some early-time exponential growth in Fig.~\ref{fig:linear}~{[Top Left]}. The fast scrambling conjecture predicts that local information is distributed over an $N$-body system within $t_*\propto\log(N)$~ \cite{2008JHEP...10..065S}. In fact, comparing $t_*$ read from Fig. 5 of \cite{2020PhRvL.125m0601B}, I can say that  scrambling happens faster in this model. Moreover according to~\cite{2020PhRvL.125m0601B}, a necessary condition for fast scrambling is that, before the onset of exponential growth, the decay of correlations with $N$ should be at most algebraic $C\propto N^{-\alpha}$ and not exponential. Fig.~\ref{fig:linear}~[Top Right] shows $C\propto N^{-3}$ between the two ends of the chain after a fixed time. Furthermore the entanglement grows non-linearly in time before saturating and shows a significant speed up as exhibited in Fig.~\ref{fig:linear}~[Bottom].\if{ Moreover Fig.~\ref{fig:MI} shows the negativity of the tripartite information $I_3(A:C:D)$, which is a sign of scrambling~\cite{2016JHEP...02..004H}. }\fi~Therefore the model I studied in this Comment satisfies all criteria of fast scrambling. It has only short-range interactions, and the number of interactions and the dimension of the Hilbert space are smaller than those of~\cite{2020PhRvL.125m0601B}, which the authors called "minimal". In short the "minimal" model \cite{2020PhRvL.125m0601B} is not the minimal in the literature. 

I thank Viktor Jahnke for reading the manuscript carefully and giving me appropriate feedback.  

\bibliographystyle{utphys}
\bibliography{ref}

\providecommand{\href}[2]{#2}\begingroup\raggedright\begin{thebibliography}{1}

\bibitem{2020PhRvL.125m0601B}
R.~{Belyansky}, P.~{Bienias}, Y.~A. {Kharkov}, A.~V. {Gorshkov}, and
  B.~{Swingle}, ``{Minimal Model for Fast Scrambling},''
  \href{http://dx.doi.org/10.1103/PhysRevLett.125.130601}{{\em \prl} {\bfseries
  125} no.~13, (2020) 130601}.

\bibitem{2008JHEP...10..065S}
Y.~{Sekino} and L.~{Susskind}, ``{Fast scramblers},''
  \href{http://dx.doi.org/10.1088/1126-6708/2008/10/065}{{\em Journal of High
  Energy Physics} {\bfseries 2008} no.~10, (2008) 065}.

\end{thebibliography}\endgroup

\end{document}